\documentclass[journal=jacsat,manuscript=article]{achemso}
\usepackage[utf8]{inputenc}
\usepackage{textcomp}
\usepackage{graphicx}
\usepackage[version=4]{mhchem}
\usepackage{siunitx}
\DeclareSIUnit\au{a.u.}
\usepackage{comment}
\usepackage{epstopdf}
\AppendGraphicsExtensions{.tiff}
\usepackage{gensymb}
\usepackage{tabularx}
\usepackage{multirow}

\title{Ab Initio Investigation of Intramolecular Charge Transfer States in DMABN by Calculation of Excited State X-ray Absorption Spectra}

\author{Avdhoot Datar}
\email{adatar@smu.edu}
\affiliation{Department of Chemistry, Southern Methodist University, Dallas, TX 75275, USA}
\altaffiliation{These authors contributed equally.}

\author{Saisrinivas Gudivada}
\affiliation{Department of Chemistry, Southern Methodist University, Dallas, TX 75275, USA}
\alsoaffiliation{Present Address: Department of Physics, The University of California at Berkeley, Berkeley, CA 94720, USA}
\altaffiliation{These authors contributed equally.}

\author{Devin A. Matthews}
\email{damatthews@smu.edu}
\affiliation{Department of Chemistry, Southern Methodist University, Dallas, TX 75275, USA}

\abbreviations{ES-XAS, DMABN, ICT, PICT, TICT, WICT}
\keywords{Transient X-ray Absorption, Intramolecular Charge Transfer, Dual Fluorescence, coupled cluster, equation-of-motion}

\begin{document}

\begin{tocentry}
\includegraphics[scale=0.48]{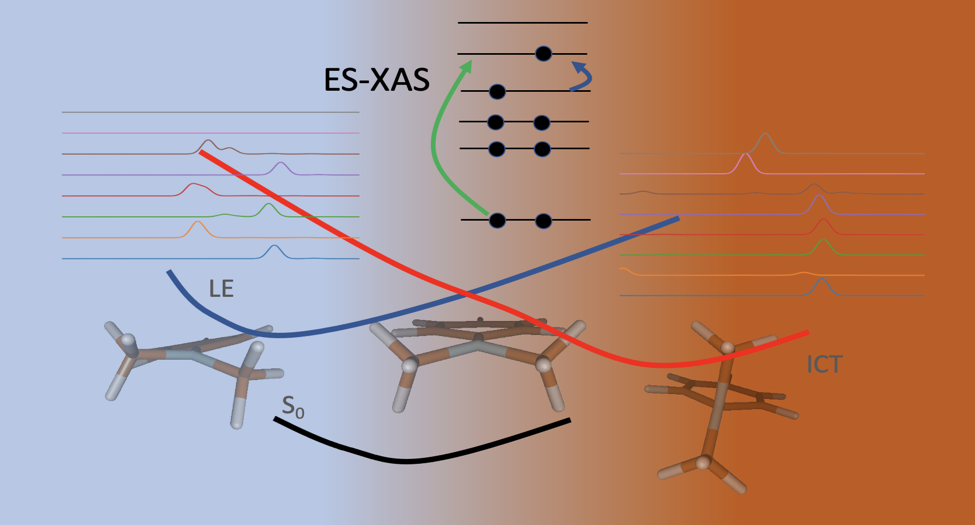}  
\end{tocentry}

\begin{abstract}
Dual fluorescence in 4-(dimethylamino)benzonitrile (DMABN) and its derivatives in polar solvents has been studied extensively for the past several decades. An intramolecular charge transfer (ICT) minimum on the excited state potential energy surface, in addition to the localized low-energy (LE) minimum, has been proposed as a mechanism for this dual fluorescence, with large geometric relaxation and molecular orbital reorganization a key feature of the ICT pathway. Herein, we have used both equation-of-motion coupled-cluster with single and double excitations (EOM-CCSD) and time-dependent density functional (TDDFT) methods to investigate the landscape of excited state potential energy surfaces across a number of geometric conformations proposed as ICT structures. In order to correlate these geometries and valence excited states in terms of potential experimental observables, we have calculated the nitrogen K-edge ground and excited state absorption spectra for each of the predicted ``signpost" structures, and identified several key spectral features which could be used to interpret a future time-resolved x-ray absorption experiment.
\end{abstract}

\section{Introduction}
\begin{figure}
\includegraphics[scale=0.4]{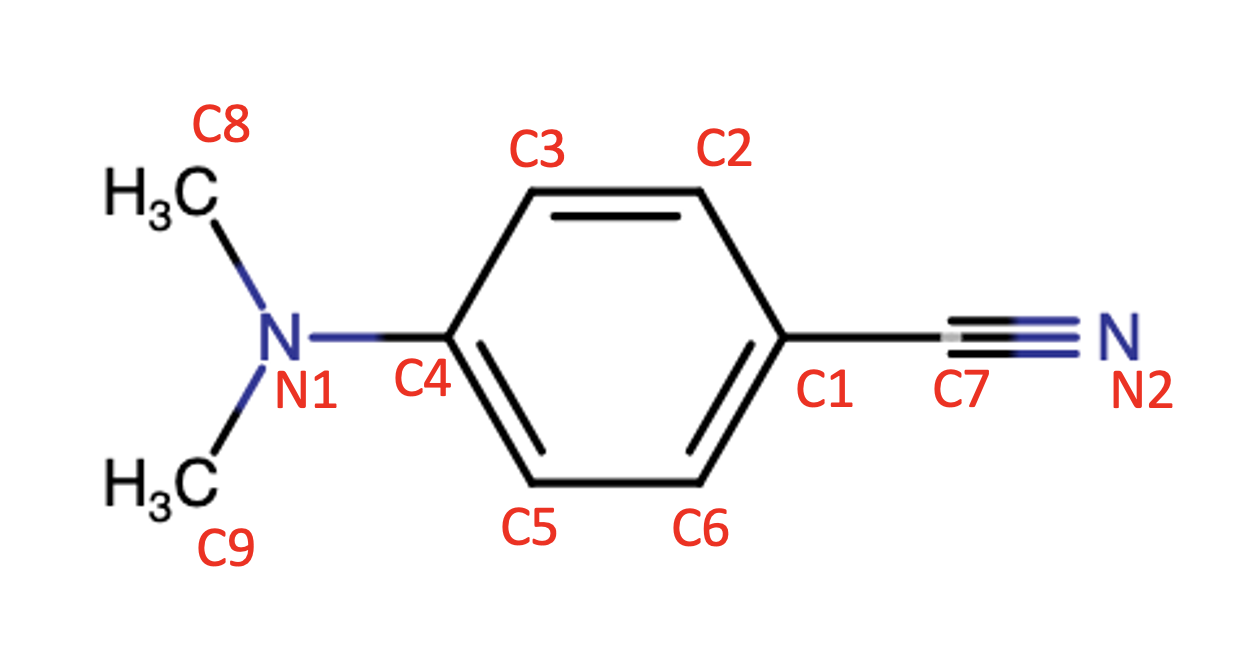}\caption{Schematic structure of DMABN with non-hydrogen atoms labelled with numbers. \label{fig:schematic}}
\end{figure}

Since the discovery of dual fluorescence in 4-(dimethylamino)benzonitrile (DMABN, Fig.~\ref{fig:schematic}),\cite{lippert1961umwandlung} the valence excited electronic states of DMABN and their evolution following photoexcitation have been studied extensively to gain mechanistic understanding of this phenomenon.\cite{modesto2018discrete,kochman2020simulating,curchod2017ab,sobolewski1998ab,parusel1999coupled,rappoport2004photoinduced,kohn2004,gomez2021micro} Initially the state reversal model was proposed to explain the dual fluorescence.\cite{lippert1962advances} In the context of this model, it was suggested that initial excitation to the lowest absorption band leads to population of S$_1$ and S$_2$ states. Internal conversion from the S$_2$ state followed by intramolecular vibrational relaxation (IVR) on the S$_1$ potential energy surface (PES) leads to ``normal" fluorescence from the S$_1$ minimum. Alternatively, the ``anomalous" fluorescence, which is sensitive to solvent polarity and temperature, was attributed to the S$_2$ minimum. Based on the nature of the orbitals taking part in these electronic excitations, the S$_1$ state is termed the locally excited (LE) state, while the S$_2$ state is classified as an intramolecular charge transfer (ICT) state. Grabowski et. al. suggested a mechanism that involved, on the ICT PES, twisting of the amino group relative to the aryl ring reaching an angle of 90\degree.\cite{rotkiewicz1973reinterpretation,grabowski1979twisted} This proposed minimum on the ICT PES is denoted the twisted ICT (TICT) structure. Later, it was shown that this TICT state cannot be populated via direction absorption.\cite{leinhos1991intramolecular}

Various models that have been proposed to explain the anomalous fluorescence mainly differ in the minimum energy structure assigned to the ICT state.\cite{grabowski2003structural} In contrast the the model outlined above, the wagging motion of the dimethylamino group was also suggested as a primary coupling mode of the LE and ICT states, which leads to a proposed wagged ICT (WICT) structure.\cite{zachariasse1996intramolecular,zachariasse1997photo} A model based on a planar minimum (with respect to the amino pyramidalization angle) was also put forth giving a planar ICT (PICT) state.\cite{zachariasse2000comment,zachariasse2002picosecond,zachariasse2004intramolecular} Additionally, a bending on the cynano group within the plane of the aryl ring, in principle produced by a rehybridization of the cyano nitrogen orbitals from sp to sp$^2$, leads to a rehybridized ICT (RICT) hypothesis.\cite{lewis1980singlet,sobolewski1996promotion} Lastly, a partially-twisted ICT geometry, as opposed to the full $90^\degree$ twist angle of TICT, is suggested to be responsible for driving population of the ICT state from the wagged ground state geometry, giving a partially-twisted ICT (PTICT) state.\cite{coto2011intramolecular} Each of these models invoke only LE and ICT states and their associated PESs as a mechanism for explaining dual fluorescence; we also mention for completeness another study where it was proposed that a third $\pi\sigma^*$ state takes part in the mechanism along with LE and ICT states.\cite{georgieva2015intramolecular}

Although the TICT model is the most well-established mechanism of dual fluorescence in DMABN, vigorous ongoing debate fuels the investigation of other models as well as efforts to probe the finer details of the mechanism. Despite various theoretical approaches to understand the fluorescence mechanism, there exists no unambiguous model that can explain this phenomenon, and a cohesive description based on simple models is elusive.\cite{rappoport2004photoinduced} As discussed by K{\"o}hn and H{\"a}ttig, a central goal of the theoretical study on this problem is to test various models for consistency and to offer a prediction that can be investigated in experimental studies.\cite{kohn2004} 

In this work we follow the spirit of K{\"o}hn and H{\"a}ttig and compute the excited state x-ray absorption spectra (ES-XAS) of multiple valence excited states at several geometries relevant to potential dual-fluorescence pathways.
X-ray absorption spectroscopy (XAS), as it involves excitation of core-level electrons to the unoccupied valence/Rydberg orbitals, is an effective technique to study the local electronic as well as geometrical structure. Another attractive aspect of XAS is that it provides atomic specificity. Thus, high-resolution XAS is widely applied as a powerful technique for determining the local structure of molecular systems.\cite{van2016x,bunker2010introduction,Stohr1992,mcneilXrayFreeelectronLasers2010,garg1994x} While XAS provides a useful static picture, time-resolved XAS (tr-XAS) spectroscopy is employed for tracking dynamical processes, such as solute-solvent interactions, evolution of photoexcited states, intramolecular charge transfer processes, etc. With recent advancements in experimental techniques, tr-XAS is regularly used to study condensed phase systems.\cite{balerna2015synchrotron,bressler2010molecular,costantini2019picosecond} Our calculations of the nitrogen K-edge (N1s) ES-XAS of DMABN will inform potential experimental studies by providing detailed, geometry- and state-dependent excited state XAS features and their interpretation in terms of molecular orbitals via natural transition orbitals (NTO) analysis.

\section{Computational Details}

The ground and excited state geometries were optimized using coupled cluster with single and double excitations (CCSD)\cite{purvisFullCoupledCluster1982,scuseriaClosedshellCoupledCluster1987,leeEfficientClosedshellSingles1988} and equation-of-motion CCSD (EOM-CCSD)\cite{sekinoLinearResponseCoupledcluster1984,geertsenEquationofmotionCoupledclusterMethod1989,stanton1993equation} methods, respectively. The EOM-CCSD method was also used to optimize the conical intersection (CI) geometry along the intersection seam.\cite{KOGA1985371} X-ray absorption spectra were simulated using the core-valence-separated equation-of-motion coupled cluster with single and double excitations (CVS-EOM-CCSD).\cite{corianiCommunication2015} All XAS spectra were broadened using a Gaussian profile with a standard deviation of \SI{0.2}{eV}. The Pople split-valence triple-$\zeta$ basis set with diffuse and polarization functions, 6-311++G**,\cite{krishnan1980self,clark1983efficient} was used for all calculations, unless specifically noted. Optimizations and (ES-)XAS calculations were performed with a development version of the CFOUR software package.\cite{Matthews2020} Natural transition orbitals (NTOs) were used to identify the nature of the excited states.

To better understand the effects of geometry on the ES-XAS, molecular geometries were also optimized using DFT methods. The ground state geometry was obtained with the M06-2X exchange-correlation functional.\cite{Zhao2008} The excited state and CI geometry optimizations were performed using the BHHLYP exchange-correlation functional\cite{becke1993new} and the Tamm-Dancoff approximation of time-dependent density functional theory (TDDFT). These calculations were performed using the Q-Chem software package.\cite{doi:10.1080/00268976.2014.952696}
Valence and core excitation energies, oscillator strengths, and dipole moments for these geometries were then calculated using EOM-CCSD and CVS-EOM-CCSD methods as above.

\section{Results and Discussion}

\subsection{Ground state structure}

\begin{figure}
\includegraphics[scale=0.45]{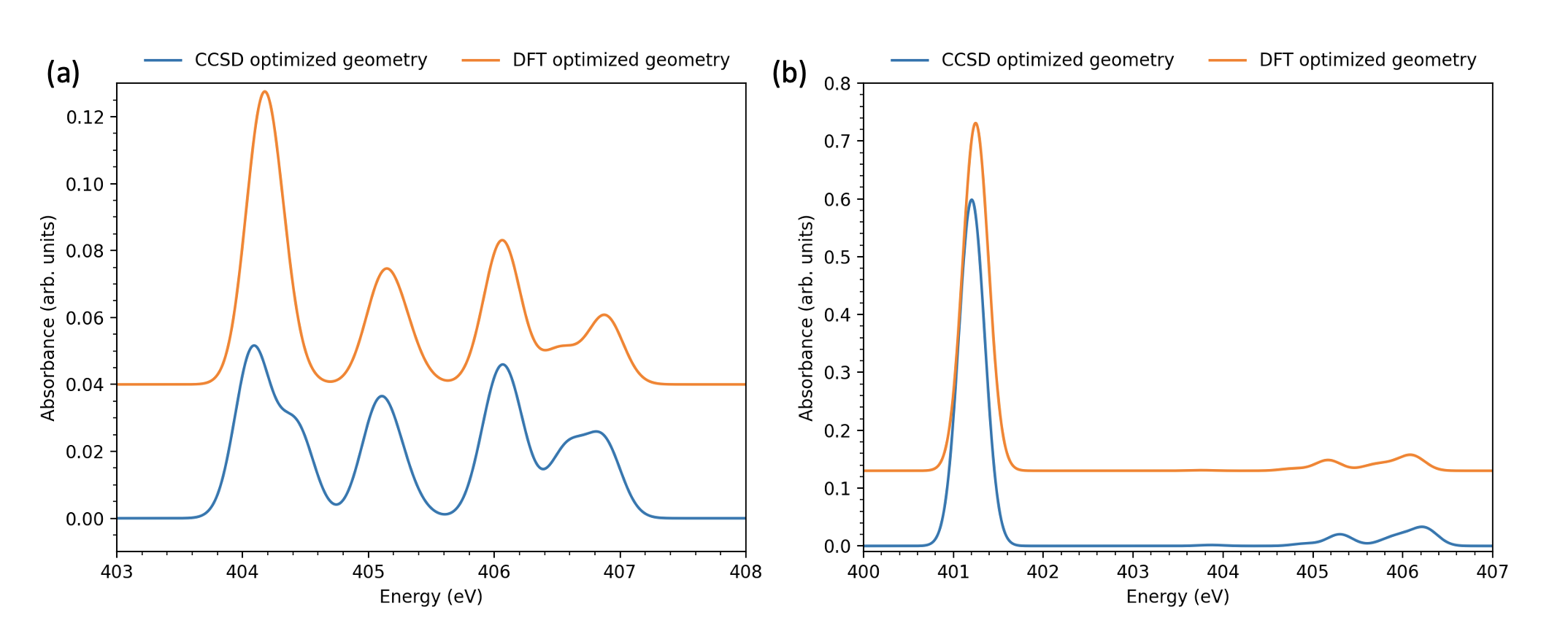}\caption{XAS calculation for the ground state geometries obtained from the CCSD and the DFT calculations: N1s excitation of (a) amino and (b) cyano nitrogen atoms. \label{fig:XAS-ground}}
\end{figure}

A slight pyramidalization of the dimethylamino group of DMABN has been established from microwave gas phase spectroscopy\cite{kajimoto1991structure} and from x-ray diffraction in the crystalline phase\cite{heine1994structure}. The wagging distortion with respect to the planar structure is observed as 15\degree and 11.9\degree in gas phase and in crystalline phase, respectively. Parusel et. al. have shown that triple-$\zeta$ quality basis sets are necessary to obtain the ground state geometry with an adequate pyramidalization angle.\cite{parusel1998density}

Our CCSD optimization of the ground state geometry leads to a non-planar structure with distortion along the wagging motion of the dimethylamino group (\ce{\angle C3C4N1C8} $\sim$ 19\degree, numbering as in Fig.~\ref{fig:schematic}) in good agreement with the results obtained from gas phase spectroscopy.\cite{kajimoto1991structure} Alternatively, the DFT optimization leads to a mostly planar structure (\ce{\angle C3C4N1C8} $\sim$ 5.4\degree).

Fig.~\ref{fig:XAS-ground} illustrates the effect of the pyramidalization angle on the ground state XAS. Although the spectra calculated for both the geometries is identical for N1s excitation of the cyano nitrogen, and are mostly similar for the amino nitrogen, the latter XAS spectra show a distinctive splitting of the lowest energy peak at less-planar CCSD geometry. The magnitude in the splitting is only approximately \SI{0.4}{eV}, however, which is within the typical errors of relative XAS peak positions for CVS-EOM-CCSD \cite{matthewsEOMCCMethodsApproximate2020}. Nonetheless with more detailed follow-up calculations this distinctive feature could then provide an additional spectroscopic measure of the ground state pyramidalization angle.

\subsection{Valence excited states} 

\begin{table}
\caption{Excitation energies, oscillator strengths ($f$) and dipole moments ($\mu$) of valence excited states at the CCSD optimized ground state geometry.\label{valence-excited-states}}

\begin{tabular}{|m{2cm}|m{2cm}|m{2cm}|>{\centering\arraybackslash}m{2cm}|m{2cm}|m{2cm}|m{2cm}|>{\centering\arraybackslash}m{2cm}|}
\hline 
\multirow{2}{*}{State} & \multicolumn{3}{|c|}{6-311++G**} & \multicolumn{3}{|c|}{6-311G**} \\
\cline{2-7}
  & $\Delta$E (eV)  &  $f$ &  $\mu$ (D) & $\Delta$E (eV)  &  $f$ & $\mu$ (D) \\
\hline 
LE  & 4.59  & 0.0212 & 8.67 & 4.67  & 0.0195 & 8.42\\
\hline 
pre-ICT  & 4.93  & 0.3307 & 6.35 & 5.18  & 0.5751 & 12.07\\
\hline 
S$_3$  & 5.24  & 0.2604 & 7.50 & 6.49  & 0.0249 & 2.52\\

\hline 
\end{tabular}
\end{table}

It is well-established that the LE state is the lowest excited singlet state of DMABN in the Franck-Condon (FC) region, while the ICT state is the second lowest. The nature of local and charge-transfer excitations of the vertical excited states were confirmed from NTO analysis as shown in Table~S1 of the ESI. Previous theoretical calculations of these states predicted a significantly higher dipole moment of the ICT state, consistent with the assignment of charge-transfer character.\cite{sobolewski1998ab,parusel1999coupled,rappoport2004photoinduced,kohn2004} However, our calculations result in a dipole moment of the ICT state lower even than that of the LE state (Table~\ref{valence-excited-states}). The previous theoretical calculations were performed with basis sets that did not include diffuse functions, and so we recomputed the vertical excited states at the same geometry but with the non-diffuse 6-311G** basis set. The latter calculations do indeed show a significant increase in dipole moment for the ICT state, but also a drastic destabilization of the third singlet state S$_3$ (Table~\ref{valence-excited-states}). Without diffuse functions, this state is clearly identified as a Rydberg state (Table~S1), while the inclusion of diffuse functions leads to stabilization and mixing with the ICT state. This mixing is then responsible for the dilution of charge-transfer character and a lowering of the dipole moment. This comparison suggests that while the LE state is relatively stable with respect to the quality of the basis set used, the ICT state is highly sensitive to the basis set, especially the inclusion of diffuse functions. Due to this Rydberg mixing, which is not significant at other geometries (vide infra), we denote the ICT state in the Frank-Condon region as a ``pre-ICT" state. We also find a strong dependence of the ground- and excited state-XAS on diffuse functions, as illustrated in Figs.~S1--S3. The DFT optimized ground state, which is nearly planar, also leads to a large dipole moment of the ICT/pre-ICT state due to a reduction in Rydberg mixing. Thus, the use of a sufficiently flexible and diffuse basis set with an accurate electronic structure method seems critical for a correct qualitative or quantitative description of the excited states of DMABN.

\begin{table}
\caption{Dipole moments (in Debye) of LE and ICT states for various optimized geometries calculated at the EOM-CCSD/6-311++G** level of theory.\label{dipole-moments}}

\begin{tabular}{|m{2cm}|m{2cm}|m{2cm}|>{\centering\arraybackslash}m{2cm}|m{2cm}|m{2cm}|>{\centering\arraybackslash}m{2cm}|}
\hline 
\multirow{2}{*}{Geometry} & \multicolumn{3}{|c|}{EOM-CCSD optimized} & \multicolumn{3}{|c|}{TDDFT optimized} \\
\cline{2-7}
  & Structure type  & LE &  ICT & Structure type  & LE  & ICT\\
\hline 
FC  & Wagged  & 8.67 & 6.35 & Wagged/ planar  & 9.40 & 12.43\\
\hline 
LE$_{min}$  & Partially twisted  & 9.66 & 12.64 & Partially twisted  & 9.67 & 9.91\\
\hline 
ICT$_{min}$  & Twisted  & 15.57 & 15.73 & Twisted  & 15.10 & 14.71\\
\hline 
S${2}$/S${1}$  & Wagged with ring distortion  & 8.77 & 8.66 & Wagged  & 9.16 & 5.53\\
\hline
\end{tabular}
\end{table}

Optimization of the LE state starting with the FC geometry leads directly to a minimum-energy structure denoted LE$_{min}$. In this structure we observe a twisting of the dimethylamino group of about 18.5\degree relative to the aryl ring. TDDFT optimization of LE results in a similar geometry but with a larger torsion angle of 30.5\degree. The torsion angle for LE$_{min}$ obtained from EOM-CCSD optimization is in a good agreement with previous results.\cite{kohn2004} A geometry optimization of the pre-ICT state starting at the FC geometry leads to a conical intersection seam (S$_2$/S$_1$). We further performed a minimum energy crossing point (MECP) search, which led to a wagged geometry, albeit with a smaller wag angle compared to the ground state geometry and a distortion of the aryl ring. Similar results were obtained from the TDDFT calculations (Table \ref{dipole-moments}). We found a minimum of the ICT state (ICT$_{min}$) at a fully-twisted geometry, and were unable to find a charge-transfer minimum energy structure on the S$_1$ surface at any partially-twisted geometry. We were also unable to find any minimum energy structure corresponding to a rehybridized (RICT) state. It is notable that while both the LE$_{min}$ and ICT$_{min}$ structures feature a twisting of the dimethylamino group, the conical intersection which, at least according to our computed potential energy surfaces, should drive the dual fluorescence mechanism via population splitting, is found at a non-twisted geometry. In the next section we illustrate the ramifications of the structural parameters of each of these geometries on the ES-XAS spectra.

The dipole moments of both LE and ICT states were found to increase with increasing twisting angle for the EOM-CCSD optimized structures (Table \ref{dipole-moments}), although NTO analysis confirms local excitation character of S$_1$ at LE$_{min}$ and charge-transfer character at ICT$_{min}$ (Tables~S3 and S4). This observation is very similar to previous calculations performed with the RI-CC2 method.\cite{kohn2004} On the other hand, previous calculations performed with CASSCF (on CIS optimized structures) predicted that the dipole moment of the LE state should decrease with increasing torsion angle.\cite{sobolewski1998ab} The dipole moment of both states also increases slightly at the CI, although Rydberg mixing is still evident. Interestingly, the CI geometry is found at a larger wag angle than the ground state using TDDFT, which leads to a reduction in the ICT dipole moment.

\subsection{Excited state XAS} 

ES-XAS spectra were calculated at the FC, LE$_{min}$, ICT$_{min}$, and S$_2$/S$_1$ geometries. The N1s core-excited states were calculated for both the amino and cyano nitrogen atoms. Transition properties such as oscillator strengths were computed with respect to the valence states S$_1$ and S$_2$ calculated as above. The spectra are labeled as ``LE" or ``ICT" ES-XAS, which assignment depends on the geometry---for example ICT is S$_1$ at ICT$_{min}$ while it is S$_2$ elsewhere.

\begin{figure}
\includegraphics[scale=0.4]{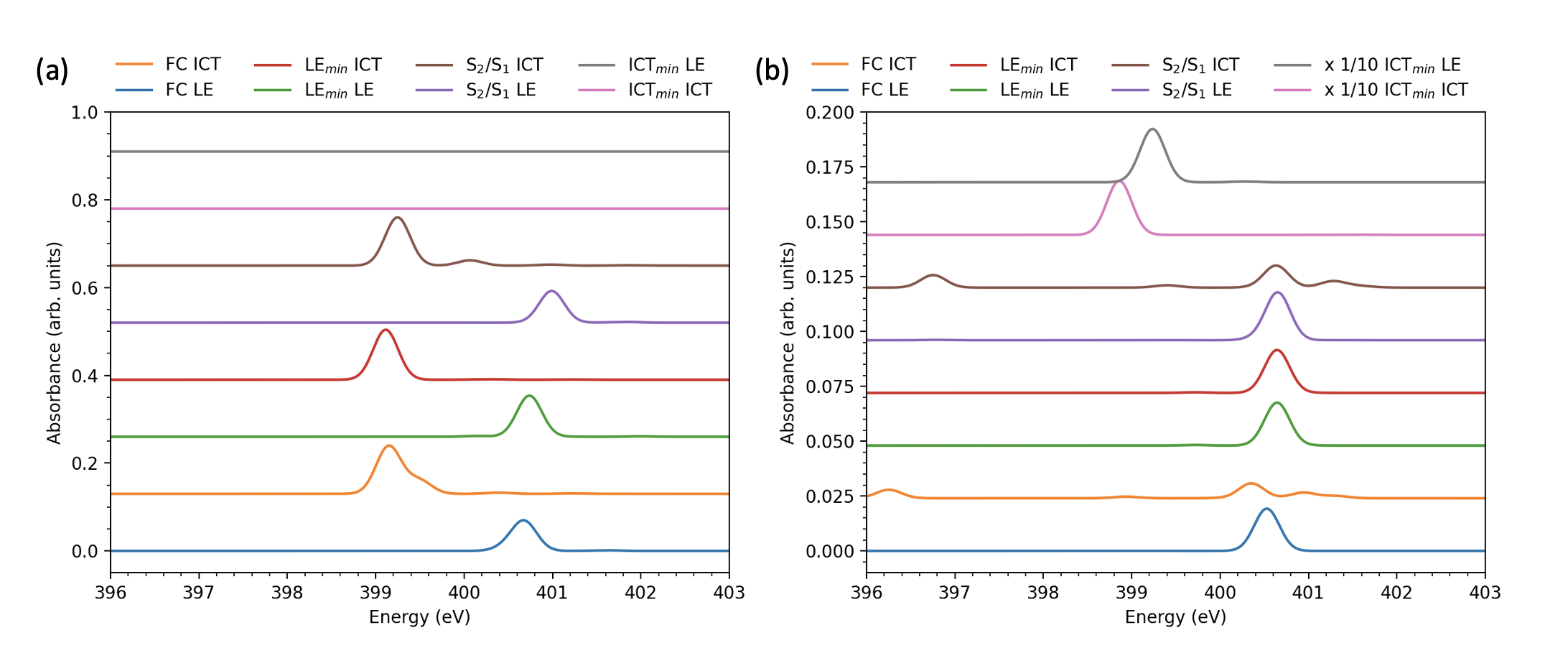}\caption{ES-XAS calculation for the geometries obtained from the CCSD calculations: N1s excitation of (a) amino and (b) cyano nitrogen atoms. \label{fig:ES-XAS-CCSD}}
\end{figure}

\begin{figure}
\includegraphics[scale=0.4]{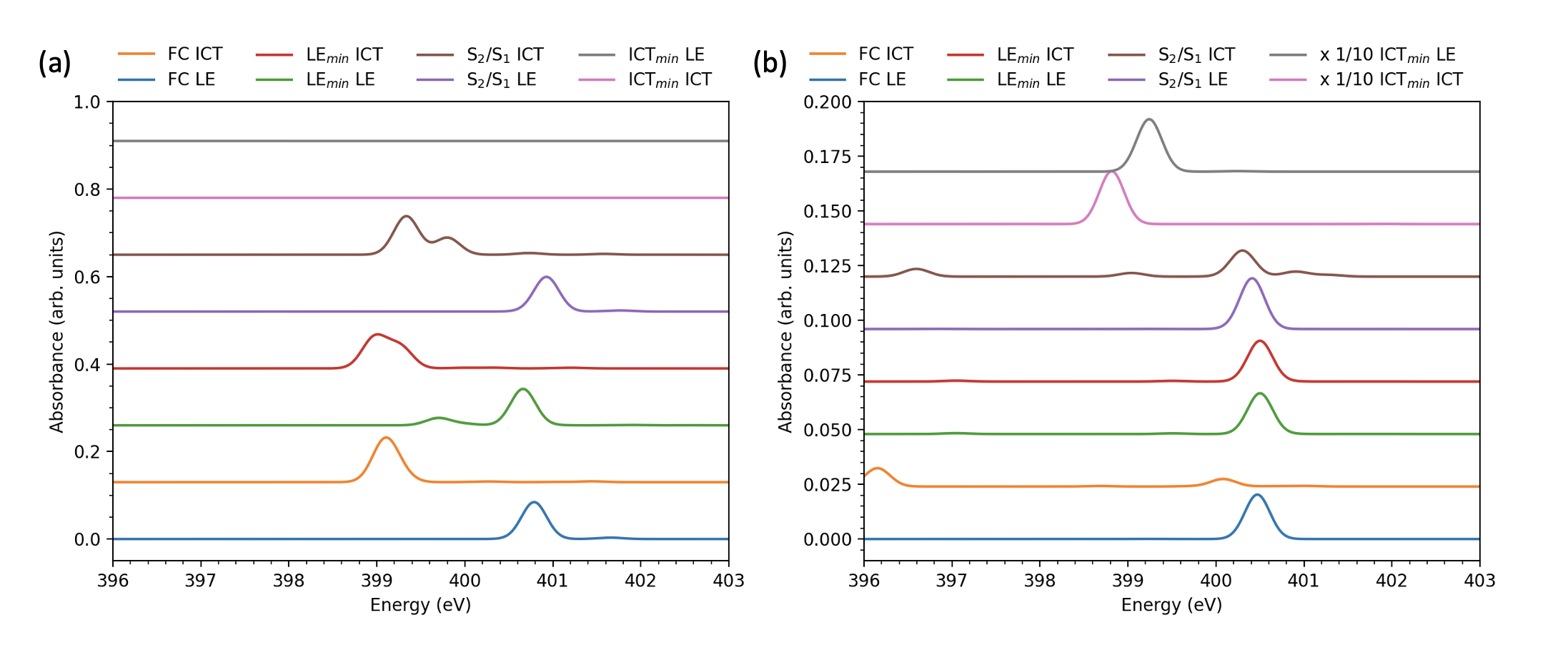}\caption{ES-XAS calculation for the geometries obtained from the DFT calculations: N1s excitation of (a) amino and (b) cyano nitrogen atoms. \label{fig:ES-XAS-DFT}}
\end{figure}

Fig.~\ref{fig:ES-XAS-CCSD} shows the ES-XAS spectra for the geometries obtained from the CCSD optimizations. The amino N1s ES-XAS show strong dependence on the lower (valence) excited state and the pyramidal distortion of the dimethylamino group. A strong peak near \SI{399}{eV} appears for the (pre-)ICT state, while for the excitation from the LE state a peak appears near \SI{400.8}{eV}, blue-shifting to \SI{\sim 401}{eV} at the IC geometry. This differential signal should provide for an unambiguous identification of the relative S$_1$ and S$_2$ populations. For the fully-twisted ICT$_{min}$ geometry, there is no ES-XAS signal at the amino nitrogen. Instead, a high-intensity peak appears in the cyano nitrogen N1s spectrum for both LE and ICT states (note the scaling factor in Fig.~\ref{fig:ES-XAS-CCSD}. This high intensity results from electron density accumulation in the cyano $\pi^*$ orbital, which also results in a slight lengthening of the CN bond length ($\sim 0.5\%$). These features clearly identify the ICT minimum structure in a potential tr-XAS spectrum. The cyano N1s spectra are otherwise weakly dependent on the state and geometry, with the exception of the ICT spectrum at wagged (non-twisted) geometries. There, the intensity of the peak near \SI{400.5}{eV} is diminished while an additional weak peak around 396--\SI{396}{eV} appears. These features might also identify the transition from a wagged to twisted geometry during the dual fluorescence process. Note that the amino and cyano absorption features near \SI{400.5}{eV} do overlap (the observable XAS is the sum of the amino and cyano spectra). However, the most interesting and sensitive features occur outside of this range, and with clearly different intensities.

The ES-XAS spectra calculated at the DFT optimized structures show strikingly similar behavior (Fig.~\ref{fig:ES-XAS-DFT}). Note that the nearly-planar ground state structure obtained from the DFT optimization shows a prominent single peak near \SI{399}{eV} (Fig.~\ref{fig:ES-XAS-DFT}a) for amino N1s excitation from the ICT state, in contrast to the shouldered peak in Fig.~\ref{fig:ES-XAS-CCSD}. Instead, the same peak at the LE$_{min}$ and IC geometries exhibits a shoulder or complete splitting. As stated earlier, the LE$_{min}$ geometry obtained from EOM-CCSD has a smaller twisting angle than in the TDDFT optimized structure. While the fine details of the spectra differ between the EOM-CCSD and TDDFT optimized structures, the robustness of the key spectral features is encouraging for their use in assigning potential tr-XAS spectra of the DMABN dual fluorescence mechanism.

\section{Conclusions}

We computed minimum energy structures on the ground, LE and ICT potential energy surfaces as well as the minimum energy point on the S$_2$/S$_1$ conical intersection seam using EOM-CCSD and TDDFT methods. These optimizations show overall agreement in the structural types (for example partially-twisted, twisted, wagged, etc.) for respective states, except the detailed geometrical parameters (wagging and torsional angles). The sensitivity of the pre-ICT state dipole moment at the ground state geometry to the diffuseness of the basis set hints that additional in-depth calculations of the ICT PES are required with larger basis sets and accurate electronic structure methods. The present calculations also show that some higher valence electronic states could also be important in the dual-fluorescence mechanism due to valence-Rydberg mixing.

We also computed ground and excited state x-ray absorption spectra for the lowest two valence states at various optimized structures on the ground and excited PESs. Several distinctive features are present which could be used to distinguish both the initial valence state involved in the absorption and the nature of the instantaneous nuclear geometry. We propose that the predicted spectral features could be used as spectroscopic ``signposts" to assign a potential time-resolved XAS spectrum, and to distinguish between LE and ICT fluorescence pathways in real time.

\begin{acknowledgement}
All calculations were performed on the ManeFrame II computing system at SMU. This work was supported by the Robert A. Welch foundation under grant no. N-2072-20210327 and in part by the National Science Foundation (awards OAC-2003931 and CHE-2143725).
\end{acknowledgement}

\begin{suppinfo}
Electronic supplementary information files are available free of charge at the publisher's website. The following files are included:
\begin{itemize}
\item Supporting\_Information.docx: NTOs for valence states at all discussed geometries, XAS and ES-XAS for non-diffuse basis sets.
\item Supporting\_Information\_Raw\_Data.xlsx: optimized geometries, valence-excited excitation energies and corresponding oscillator strengths, and \
core-excited excitation energies and corresponding oscillator strengths.
\end{itemize}
\end{suppinfo}

\bibliography{achemso-demo}

\end{document}